# Structural, Elastic, Electronic, and Optical Properties of Layered TiN$X$ ($X$ = F, Cl, Br, I) Compounds: a Density Functional Theory Study

M.M. Hossain[1] and S.H. Naqib[2*]

[1]Industrial Physics Division, BCSIR Laboratories Dhaka, BCSIR, Dhaka 1205, Bangladesh
[2]Department of Physics, University of Rajshahi, Rajshahi 6205, Bangladesh

*Corresponding author; Email: salehnaqib@yahoo.com

**Abstract**
Titanium nitride halides, TiN$X$ ($X$ = F, Cl, Br, I) in the $\alpha$-phase (orthorhombic) are exciting quasi two-dimensional (2D) electronic systems exhibiting a fascinating series of electronic ground states under different conditions. Pristine TiN$X$ are semiconductors with varying energy gaps and possess attractive properties for potential applications in the fields of optoelectronics, photovoltaics, and thermoelectrics. Alkali metal intercalated TiNCl becomes superconducting at reasonably high temperature in the α-phase. We have revisited the electronic band structure of these compounds using density functional theory (DFT) based first-principles calculations. The atomic species and orbital resolved partial electronic energy density of states are calculated together with the total density of states (TDOS). The structural and elastic properties have been investigated in details for these layered compounds for the first time. The elastic anisotropy has been explored. The optical properties, including energy dependent real and imaginary parts of the dielectric constant, optical conductivity, reflectivity, and loss function of TiN$X$ are studied for the first time. The Debye temperatures of these compounds have been calculated and the related thermal and phonon parameters are discussed. The calculated physical parameters are compared with existing theoretical and experimental results and show fair agreement, where available. All these compounds are found to reflect electromagnetic radiation strongly in the mid ultraviolet region. The elastic properties show high degree of anisotropy. The lattice is highly compressible along the crystallographic $c$-direction. The effect of halogen atoms on various structural, elastic, electronic, and thermal properties in TiN$X$ are also discussed in detail.

# 1 Introduction

Layered TiN$X$ ($X$ = F, Cl, Br, I) compounds belong to a series of transition metal nitride halides with many fascinating physical properties [1 – 3]. The general formula of these layered-structured compounds are *M*N*X*, where *M* = Ti, Zr, Hf, the group 4 transition elements and *X* = F, Cl, Br, I, the halogen elements. A significant number of researches involving *M*N*X* were focused on superconductivity [1 – 7]. These layered compounds exhibit superconductivity with reasonably high transition temperatures, when intercalated by alkali metals (Li, Na, K, Rb) [7]. For example, HfNCl intercalated by alkali metal is an electron doped metal and becomes superconducting at ~ 25.5 K [8]. These transition metal nitride halide compounds have easily tunable electrical transport properties, and their layered structure often facilitates to introduce additional phonon scattering channels via interlayer atomic interactions, which tend to cause the materials to exhibit intrinsically low thermal conductivity [9]. This makes *M*N*X* compounds as promising candidates for thermoelectric device applications [10].

*M*N*X* compounds are found in two different types of structure. The first type is the FeOCl structure *α*-type (orthorhombic) and the second type is the SmSI structure *β*-type (hexagonal) [11]. The most widely studied intercalated superconductor metal nitride halide *β*-ZrNCl has a superconducting transition temperature $T_c$ = 13−15 K [6]. The *β*-type structure contains dual honeycomb layers with alternating *M* and N atomic species. TiNCl, which assumes the *α*-type structure, exhibits superconductivity at 16 K once intercalated with alkali dopants [6, 7]. In orthorhombic phase, *M*N*X* consists of *M*-N layer net which is topologically equivalent to a single NaCl layer. There is strong buckling of this layer [11] perpendicular to the *b* axis of the crystal. Due to this buckling, the neighboring chains of *M*-N atoms running along the *a*-direction differ in height [11]. The *M* ions are twofold coordinated by the halogen ions residing in the *bc*-plane. Each transition metal ion (*M*) is sixfold coordinated by N ions and two halogen ions in the structure. It should be kept into the mind that although *M*N*X* crystallizes both with the orthorhombic and hexagonal symmetries, there are significant similarities. For example both the crystal systems contain metal-nitrogen networks. The local bonding features of *α*-type and *β*-type compounds are also quite similar [4 – 6], giving rise to dispersive conduction band features with dominant *d* orbital character. It is interesting to note that, in addition to halogen atoms, *M*N*X* (e.g., TiNCl) can become superconducting via intercalation of ethylenediamine (EDA) and hexamethylenediamine (HDA) [12]. The basic mechanism relies mainly on the expansion of the distance between the basal planes from its pristine value and extent of electron doping in *M*N*X*.

Recently Zhang et al. [10] have theoretically studied the charge and thermal transport properties of TiNBr and found an ultrahigh value of the Seebeck coefficient of 2215 μV/K. The theoretically estimated thermoelectric figure of merit, *ZT*, was 0.661 at 800 K, which translated to a lattice thermal conductivity of 1.34 W/m-K. This low thermal conductivity was linked to the low value of the group velocity of the phonon modes in TiNBr and to large phonon anharmonicity. These findings reinforced the idea that TiN*X* has the potential to serve efficiently in the field of thermoelectric device applications. Very recently, Liang et al. [1] have proposed that TiN*X* can serve as highly efficient donor material in excitonic solar cells (XSCs). From their theoretical investigations Liang and coworkers [1] found that the direct band gap with moderate size, ultra-high photoresponsivity, small carrier effective mass, and low exciton binding energy render TiN*X* monolayers as promising candidates for optoelectronic and photovoltaic device applications. Moreover, TiN with different halogen terminations exhibits different HOMO and LUMO energies, allowing one to fabricate a TiN*X* hetero-bilayer that serves as a type-II donor–acceptor XSCs interface. It was demonstrated that [1] these XSCs possess high photovoltaic power conversion efficiency, with values of 18% for TiNF/TiNBr XSC, 19% for TiNCl/TiNBr XSC, and 22% for TiNF/TiNCl XSC, which are far superior to the typical layered (two dimensional) photovoltaic cell systems.

Considering the versatile physical properties of *M*N*X* compounds and TiN*X* materials in particular, it is quite surprising that a comprehensive theoretical study of the mechanical properties of these materials are still lacking. To the best of our knowledge, a detailed study of energy dependent optical parameters (real and imaginary parts of the dielectric constant, reflectivity, absorption coefficient, real and imaginary parts of the refractive index, loss function, and photoconductivity) have not been investigated yet. The Debye temperature, one of the fundamental parameters needed to study a diverse range of physical properties of crystalline solids has not been investigated for TiN*X*, as far as we are aware of. The electronic band structure and density of states have been studied before [4, 13]. Study of energy dependent optical constants complements the band structure calculations. Previous investigations of layered metallic ternaries have revealed attractive optical characteristics for possible optoelectronic device applications in several systems [14 – 18]. Therefore, study of optical properties of TiN*X* is important from both the point of view of fundamental physics and possible applications. Knowledge regarding mechanical properties, elastic anisotropy, mechanical failure modes and degree of machinability is important for device fabrication. Keeping these issues in mind we have investigated the mechanical and optical properties of TiN*X* for the first time in detail. We have also investigated systematically the Debye temperature of these materials, linked directly to phonon thermal conductivity, melting temperature and energy scale for electron-phonon superconductors.

The organization of this paper is as follows. In Section 2, we have briefly described the computational methodology employed in this paper. Section 3 comprises of the results of computations. We also discuss the salient features of the computed results in this section. Finally, in Section 4, important conclusions of this study are presented and discussed.

**2 Computational methodologies**

The most popular practical approach to *ab*-initio modeling of structural and electronic properties of crystalline solids is the DFT with periodic boundary conditions. In this formalism the ground state of the crystalline system is found by solving the Kohn-Sham equation [19]. For reliable estimates of the ground state physical properties of a system, the choice of exchange correlation scheme and functional is important. For metallic systems the generalized gradient approximation (GGA) is often a good starting point. GGA has a tendency of relaxing the crystal lattice and overestimating the lattice constants. For systems with high average electron density and a small deviation from the average, local density approximation (LDA) can be used. Unlike GGA, LDA contracts the lattice due to localized nature of the trial orbitals. Sometimes a combination of Hartree-Fock exchange term and density functional correlation term yields a better result. This is the hybrid computational scheme. All these approaches have their merits and limitations. In this study, we have used GGA, LDA and HSE06 (hybrid) schema as contained within the CAmbridge Serial Total Energy Package (CASTEP) [20] code designed to implement DFT based calculations to estimate the ground state properties of TiN$X$. By comparison with the experimentally measured lattice parameters, it was found that LDA gives the best estimates of the ground state electronic and structural parameters for the TiN$X$ compounds under study. Therefore, we have focused on LDA calculations in the subsequent sections. A typical example of the results of the calculations with different exchange-correlation formalism for TiNCl is shown in Table 1.

Vanderbilt-type ultra-soft pseudopotentials were used to model the electron-ion interactions [20]. This relaxes the norm-conserving criteria but at the same time produces a smooth and computation friendly pseudopotential which minimizes the computational time without compromising the accuracy appreciably. Broyden Fletcher Goldfarb Shanno (BFGS) geometry optimization [21] scheme has been employed to optimize the crystal structure for the given symmetry (*Pmmm*, space group No. 59). The following electronic orbitals have been used for Ti, N, F, Cl, Br, and I to derive the valence and the conduction band structures, respectively: Ti [$3p^6$ $3d^2$ $4s^2$], N [$2s^2$ $2p^3$], F [$2s^2$ $2p^5$], Cl [$3s^2$ $3p^5$], Br [$4s^2$ $4p^5$], and I [$5s^2$ $5p^5$]. Periodic boundary conditions are used to determine the total energies of each cell. Tolerance levels for computational convergence were set to ultrafine mode to ensure very high level of convergence. An energy cut-off of 500 eV was used for the plane wave basis set expansion. *k*-point sampling within the first Brillouin zone (BZ) for the compounds under study was carried out with 8×8×4 *k*-point grids following the Monkhorst-pack grid scheme [22]. Elastic constants were calculated by employing the 'stress-strain' method as included within the CASTEP program. The bulk modulus, $B$ and the shear modulus, $G$ were obtained from the calculated single crystal elastic constants, $C_{ij}$. The electronic band structure features are calculated using the theoretically optimized geometry of TiN$X$. All the optical parameters have been obtained by considering both interband and intraband transition probabilities. The imaginary part, $\varepsilon_2(\omega)$, of the complex dielectric function has been calculated from the matrix elements of electronic transition between occupied and unoccupied orbitals by employing the CASTEP supported formula expressed as

$$\varepsilon_2(\omega) = \frac{2e^2\pi}{\Omega\varepsilon_0} \sum_{k,v,c} \left|\left\langle \psi_k^c \left| \hat{u} \cdot \vec{r} \right| \psi_k^v \right\rangle\right|^2 \delta(E_k^c - E_k^v - E) \qquad (1)$$

In the above expression, $\Omega$ is the volume of the unit cell, $\omega$ frequency of the incident electromagnetic wave (photon), $e$ is the electronic charge, $\psi_k^c$ and $\psi_k^v$ are the conduction and valence band wave functions at a given wave-vector $k$, respectively. The delta function ensures conservation of energy and momentum during the optical transition. It is worth noting that Eqn. 1 has been written for interband transitions. This equation is equally valid for intraband optical transitions with relevant changes in the indices. The Kramers-Kronig transformations yield the real part $\varepsilon_1(\omega)$ of the dielectric function from the corresponding imaginary part $\varepsilon_2(\omega)$. Once these two parts of the energy dependent dielectric constant are known, all the optical parameters can be extracted from them [16, 23]. This procedure has been used extensively by a great deal of earlier works to reliably calculate the frequency dependent optical constants for compounds belonging to diverse classes [24 – 26].

**Table 1**
Optimized lattice parameters, cell volume, and band gap energy of TiNCl obtained with different exchange-correlation potentials

| $a$ (Å) | $b$ (Å) | $c$ (Å) | $V$ (Å$^3$) | Band gap energy (eV) | Remarks |
|---------|---------|---------|-------------|----------------------|---------|
| 3.889   | 3.206   | 7.553   | 94.219      | 0.503                | LDA[this study] |
| 3.954   | 3.272   | 8.691   | 112.481     | 0.612                | GGA[this study] |
| 3.868   | 3.221   | 6.835   | 85.184      | 0.713                | HSE06[this study] |
| 3.887   | 3.170   | 7.551   | 93.041      | ---                  | Ref.[13][theor.] |
| 3.937   | 3.258   | 7.803   | 100.087     | 0.600*               | Ref.[27][expt.**] |

\* Theoretical estimate of the band gap [28].
\*\* Experimental cell parameters.

## 3 Computational results

### 3.1 Crystal structure of TiN*X*

The schematic crystal structure of TiN*X* is shown below in Fig. 1. The optimized lattice constants and related cell volume is presented in Table 2. We have also given the lattice constants obtained by previous studies in this table.

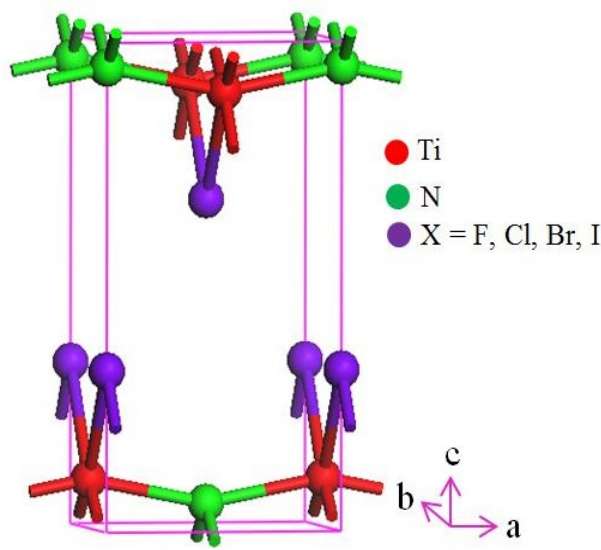

Figure 1: Schematic crystal structure of TiN$X$. The crystallographic axes directions are shown.

**Table 2**
Optimized lattice parameters and cell volumes of orthorhombic TiN$X$ ($\alpha$-phase).

| Compound | $a$ (Å) | $b$ (Å) | $c$ (Å) | $V$ (Å$^3$) | Remarks |
|---|---|---|---|---|---|
| TiNF | 3.884 | 3.020 | 5.618 | 65.890 | This study |
|  | 3.890 | 2.963 | 7.988 | 92.070 | Theoretical [13] |
| TiNCl | 3.889 | 3.206 | 7.553 | 94.219 | This study |
|  | 3.887 | 3.170 | 7.551 | 93.041 | Theoretical [13] |
|  | 3.937 | 3.258 | 7.803 | 100.087 | Expt. [27] |
| TiNBr | 3.882 | 3.306 | 8.070 | 103.570 | This study |
|  | 3.880 | 3.265 | 8.077 | 102.321 | Theoretical [13] |
|  | 3.927 | 3.349 | 8.332 | 109.578 | Expt. [27] |
| TiNI | 3.897 | 3.470 | 8.727 | 118.037 | This study |
|  | 3.885 | 3.428 | 8.702 | 115.891 | Theoretical [13] |
|  | 3.941 | 3.515 | 8.955 | 124.050 | Expt. [27] |

It seen from Table 1 that the calculated lattice parameters are in good agreement with the experimental values [27], where available. The agreement is better compared to earlier theoretical results [13]. As far as we have seen, there is no experimental data available on the lattice parameters for orthorhombic phase of TiNF. It is perhaps worth noting that experimental X-ray diffraction data are often obtained at room temperature. The theoretically optimized geometry, on the other hand, corresponds to the ground state. Therefore, the theoretical values are supposed to be somewhat lower than the experimental one because of the role played by thermal expansion. The cell volume, according to the estimates made in this study, show a systematic increase with the increase in the atomic number of $X$. This surely is primarily due to the systematic increase in the ionic radii as one moves across from F to I. There is large discrepancy in the lattice parameters and cell volume for TiNF calculated here and those calculated earlier [13]. It should be noted that F is the most electronegative among all the halogens and ionic bonding dominates in TiNF. The ionic radius of F (1.33Å) is almost 30% lower than that of Cl (1.81 Å) [29]. Therefore, very similar cell volumes found for TiNF and TiNCl in earlier estimates [13] may not be reliable. The ionic radii versus the cell volume, on the other hand, shows very nice correspondence, as far as the values obtained here are concerned.

### 3.2 *Elastic properties of* TiN*X*

The structural features, elastic response to external stress, elastic anisotropy, machinability, and bonding characteristics of crystalline solids are all intimately interlinked. TiN*X* compounds in the orthorhombic phase possess nine independent single crystal elastic constants ($C_{ij}$), given by: $C_{11}$, $C_{22}$, $C_{33}$, $C_{44}$, $C_{55}$, $C_{66}$, $C_{12}$, $C_{13}$, and $C_{23}$. We have tabulated these in Table 3 below. All these elastic constants were obtained from LDA. The elastic constants obtained by employing GGA results in significantly lower values of $C_{ij}$. This is not unexpected since for TiN*X*, GGA underestimates the atomic bonding strengths and overestimates the volume of the unit cell.

**Table 3**
Single crystal elastic constants (in GPa) of orthorhombic TiN*X* (*α*-phase).

| Compound | $C_{11}$ | $C_{22}$ | $C_{33}$ | $C_{44}$ | $C_{55}$ | $C_{66}$ | $C_{12}$ | $C_{13}$ | $C_{23}$ |
|---|---|---|---|---|---|---|---|---|---|
| TiNF  | 382.38 | 393.47 | 83.73 | 70.21 | 52.57 | 129.93 | 105.74 | 44.74 | 59.68 |
| TiNCl | 249.31 | 226.99 | 63.28 | 34.51 | 30.82 | 96.03 | 63.74 | 34.03 | 28.12 |
| TiNBr | 229.11 | 195.36 | 46.12 | 31.04 | 24.30 | 88.40 | 56.46 | 29.44 | 16.73 |
| TiNI  | 218.23 | 182.40 | 51.90 | 21.44 | 21.59 | 78.28 | 53.44 | 27.94 | 23.69 |

It is interesting to note that almost all the elastic constants decrease systematically with increasing atomic number of the halogen atom. The constants $C_{11}$, $C_{22}$, and $C_{33}$ measures the ability of the crystal to resist the applied mechanical stress along the crystallographic *a*, *b* and *c* directions respectively. It is seen that in all four compounds understudy, $C_{33}$ is much smaller than $C_{11}$ and $C_{22}$. This implies that the structure is highly compressible in the *c*-direction. This reflects the layered structure of these compounds. Bonding within the *ab*-plane is significantly stronger than those extending in the out-of-plane direction. Elastic constants $C_{44}$, $C_{55}$, and $C_{66}$ correspond to the response of the crystal to shear. These elastic constants are particularly useful because the mechanical failure modes of solids are often controlled by shearing strain, rather than the uniaxial stains. The last three elastic constants, $C_{12}$, $C_{13}$, and $C_{23}$ are due to the compounds resistance to volume conserving orthorhombic distortion.

It is possible to calculate the polycrystalline elastic moduli from the single crystal elastic constants and compliances [30, 31]. From the Hill's average values [32] of bulk (*B*) and shear (*G*) moduli, the polycrystalline Young's modulus (*Y*), and Poisson's ratio (*σ*) can be estimated [30, 31]. We have presented all these elastic parameters together with the Pugh's ratio ($B/G$) [33] in Table 4.

**Table 4**
Polycrystalline bulk moduli $B_V$, $B_R$, and $B$, shear moduli $G_V$, $G_R$, and $G$, Young's modulus $Y$ (all in GPa), Pugh's ratio $B/G$, and Poisson's ratio $\sigma$ of TiN$X$.

| Compound | $B_V$ | $B_R$ | $B$ | $G_V$ | $G_R$ | $G$ | $Y$ | $B/G$ | $\sigma$ |
|---|---|---|---|---|---|---|---|---|---|
| TiNF | 142.21 | 79.01 | 110.61 | 93.84 | 69.49 | 81.66 | 196.60 | 1.35 | 0.204 |
| TiNCl | 87.93 | 56.32 | 72.13 | 59.85 | 43.84 | 51.84 | 125.46 | 1.39 | 0.210 |
| TiNBr | 75.09 | 41.59 | 58.34 | 53.28 | 36.32 | 44.80 | 107.01 | 1.30 | 0.194 |
| TiNI | 73.63 | 46.38 | 60.00 | 47.43 | 31.89 | 39.66 | 97.50 | 1.51 | 0.229 |

In Table 4, Hill's average value of the bulk modulus was obtained from the arithmetic average of the Voight approximated [34] bulk modulus, $B_V$ and the Reuss approximated [35] bulk modulus $B_R$. Same procedure was followed to get the Hill's average value of the shear modulus from the arithmetic mean of $G_V$ and $G_R$. It is instructive to note that Voight approximation assumes a continuous strain but permits the stress to be discontinuous. As a result the actual stresses among grains are not balanced. This approximation gives an upper bound of the polycrystalline elastic moduli. The Reuss approximation, on the other hand, assumes continuous stress with discontinuous strain among the grains. As a consequence the deformed grains are not smoothly fitted with one another. This yields lower bound of the polycrystalline elastic moduli. The Hill's approximation uses the arithmetic average of these two limits and represents the real situation in the polycrystalline solids to a large extent. The bulk, shear, and the Young's modulus decrease systematically with increasing atomic number of the $X$ atomic species. Only exception is for the bulk modulus of TiNBr and TiNI, which are almost identical in magnitude. For all the compounds under study, $B > G$. This shows that TiN$X$ are prone to mechanical failure via shearing deformation. Compared to many other layered ternaries and their solid solutions, the elastic moduli of TiN$X$ are small [16 – 18, 26, 36 – 39], indicating that these materials are relatively soft. The ratio between polycrystalline bulk modulus and the shear modulus, known as Pugh's ratio [33], is a very simple and useful indicator of mechanical behavior of solids. A large value of the Pugh's ratio is associated with ductile behavior; whereas a low value implies brittleness. The brittle to ductility boundary is characterized by a critical Pugh's ratio of 1.75. It is seen that all four titanium nitride halides have Pugh's ratio well below this critical value. Therefore, we expect these materials to show brittle characteristics. Poisson's ratio is another important parameter that provides us with information not only about mechanical behavior but also about bonding characteristics. It has been shown that [40] $\sigma = 0.25$ is the lower limit for central-force solids. The low values of the Poisson's ratio of TiN$X$ indicate that interatomic forces in these solids should be non-central in nature. The abrupt brittle to ductile threshold is characterized by a Poisson's ratio of ~ 0.31 [41]. This implies that TiN$X$ are brittle in nature. A low value of Poisson's ratio for TiN$X$ also indicates that atomic packing density is low in these compounds, a characteristic of semiconducting compounds with covalent and/or ionic bonding(s). Both the Pugh's ratio and the Poisson's ratio are the highest for TiNI and these values are the lowest for TiNBr. Overall, the Pugh's ratio and the Poisson's ratio of TiN$X$ lie within narrow range 1.30 – 1.51 and 0.194 – 0.229, respectively.

Elastic anisotropy influences a number of physical processes [42] such as development of plastic deformations in crystals, propagation of cracks, and microscale cracking in ceramics, alignment or misalignment of quantum dots, enhanced mobility of charged defects, plastic relaxation of thin films, etc. Therefore, it is crucial to study the elastic anisotropy of materials for their possible engineering applications. Elastic anisotropy of crystalline solids is characterized by various anisotropy indices. An anisotropy index or factor quantify how directionally dependent the elastic properties are. In this study we have calculated a number of anisotropy factors for TiN$X$. The factors are shown in Table 5. All these anisotropy factors were calculated using widely used previously developed formalisms [30, 31, 42, 43].

**Table 5**
The shear anisotropic factors $A_1$, $A_2$, and $A_3$, anisotropy in compressibility $A_B$, anisotropy in shear $A_G$, and the universal anisotropy index $A_U$.

| Compound | $A_1$ | $A_2$ | $A_3$ | $A_B$ | $A_G$ | $A_U$ |
|---|---|---|---|---|---|---|
| TiNF | 0.746 | 0.588 | 0.921 | 0.286 | 0.149 | 2.551 |
| TiNCl | 0.565 | 0.527 | 1.101 | 0.219 | 0.154 | 2.387 |
| TiNBr | 0.574 | 0.467 | 1.135 | 0.287 | 0.189 | 3.140 |
| TiNI | 0.400 | 0.462 | 1.066 | 0.227 | 0.196 | 3.167 |

The shear anisotropic factors measure the degree of anisotropies in the bonding strength for atoms located in different crystal planes. $A_i$ = 1 (i = 1, 2, 3), implies completely isotropic behavior; departure from unity implies anisotropy. $A_1$ is the shear anisotropy factor for the {100} shear planes between the <011> and <010> directions, $A_2$ is the shear anisotropy factor for the {010} shear planes between the <101> and <001> directions, and $A_3$ is the shear anisotropy factor for the {001} shear planes between the <110> and <100> directions. It is observed that both $A_1$ and $A_2$ decrease systematically as one moves from TiNF to TiNI. With respect to $A_1$ and $A_2$, TiNF is the least anisotropic and TiNI is the most anisotropic. The values of $A_3$ for all the compounds are quite close to unity. Therefore, the response to shear in the {001} shear plane is quite isotropic in TiN$X$.

The difference between $B_V$ and $B_R$ as well as $G_V$ and $G_R$, according to Hill, are measures to the degree of elastic anisotropy of solids. Based on these limiting values, the anisotropy factors $A_B$ and $A_G$ are estimated from the following equations [44]:

$$A_B = \frac{B_V - B_R}{B_V + B_R} \quad (2)$$

$$A_G = \frac{G_V - G_R}{G_V + G_R} \quad (3)$$

These two factors allocate zero values for totally isotropic crystals. It is seen that the shear anisotropy index increases systematically as the atomic number of the halogen atom increases. The compressibility anisotropy index, on the other hand, does not show this systematic trend and remains confined within a narrow range, 0.219 to 0.287.

For an appropriate universal measure to quantify the elastic anisotropy of crystals, Shivakumar et al. [43] introduced an index termed as *universal anisotropy index*, which can be defined as:

$$A_U = 5\frac{G_V}{G_R} + \frac{B_V}{B_R} - 6 \geq 0 \quad (4)$$

This index possesses either zero or positive value, zero indicating absolutely isotropic nature of a crystal and positive value signifying the anisotropy level of a crystal. This index is considered universal because of its applicability to all crystal systems irrespective of the symmetry. From the calculated values of $A_U$, it is seen that TiNCl seems to possess the least amount of elastic anisotropy, while TiNI is the most elastically anisotropic.

Elastic constants can be used to check the mechanical stability of crystalline materials [45]. We have employed the modified Born stability criteria given by [46],

$$C_{ii} > 0 \,;\, C_{11}C_{22} > C_{12}^2;\, C_{11}C_{22}C_{33} + 2C_{12}C_{13}C_{23} - C_{11}C_{23}^2 - C_{22}C_{13}^2 - C_{33}C_{12}^2 > 0 \quad (5)$$

to investigate the mechanical stability. The inequalities given above give the necessary and sufficient elastic stability criteria for crystals with orthorhombic symmetry. The elastic constants given in Table 3 satisfy all the above inequalities and therefore, one can conclude that TiN$X$ are mechanically stable.

The ratio between the bulk modulus $B$ and $C_{44}$ can assess the machinability of a compound with the machinability index, $\mu_M = B/C_{44}$, as defined by Sun et al. [47]. This index gives a numerical value that designates the degree of difficulty or ease with which a particular material can be machined (cut or put in to different shapes). A higher value corresponds to better machinability. The calculated values of $\mu_M$ for TiNF, TiNCl, TiNBr, and TiNI were found to be 1.571, 2.090, 1.879, and 2.798, respectively.

## 3.3 Debye temperature of TiNX

As a fundamental parameter for solids, the Debye temperature, $\theta_D$, correlates with many important physical properties, such as heat capacity, bonding strengths, phonon thermal conductivity, vacancy formation energy, melting temperature etc. It also sets the characteristic boson energy scale which takes part in electron-phonon coupling and Cooper pairing in conventional superconductors. At low temperatures the vibrational excitations arise solely from acoustic vibrations. Hence, at low temperatures the Debye temperature calculated from elastic constants is identical to that determined from the specific heat measurements. Among several methods for calculating Debye temperature, Anderson method is simple and straightforward, which depends on average sound velocity and uses the following equation [48]:

$$\theta_D = \frac{h}{k_B} \left[ \left( \frac{3n}{4\pi} \right) \frac{N_A \rho}{M} \right]^{1/3} v_m \qquad (6)$$

where $h$ and $k_B$ are Planck's and Boltzmann's constants, respectively, $N_A$ is the Avogadro's number, $\rho$ refers to the mass density, $M$ stands for the molecular weight and $n$ is the number of atoms in a molecule. The sound wave, in a crystalline solid, propagates with an average velocity $v_m$ that can be determined from-

$$v_m = \left[ \frac{1}{3} \left( \frac{1}{v_l^3} + \frac{2}{v_t^3} \right) \right]^{-1/3} \qquad (7)$$

where $v_l$ and $v_t$ are the longitudinal and transverse sound velocities, respectively, in a crystalline solid. Using shear and bulk moduli ($G$ and $B$), these velocities can be determined using the following expressions [48]:

$$v_l = \left[ \frac{3B + 4G}{3\rho} \right]^{1/2} \qquad (8)$$

and

$$v_t = \left[ \frac{G}{\rho} \right]^{1/2} \qquad (9)$$

The calculated Debye temperature $\theta_D$ along with sound velocities $v_l$, $v_t$, and $v_m$ for the TiNX compounds are listed in Table 6. This particular method for calculating the Debye temperature from the elastic constants has been used extensively to reliably estimate $\theta_D$ for variety of compounds with different electronic ground states [14, 16, 17, 30, 49 – 51]

**Table 6**
Calculated density ($\rho$ in gm/cm$^3$), longitudinal, transverse, and average sound velocities ($v_l$, $v_t$, and $v_m$ in km/s) and Debye temperature ($\theta_D$ in K).

| Compound | $\rho$ | $v_l$ | $v_t$ | $v_m$ | $\theta_D$ |
|---|---|---|---|---|---|
| TiNF | 4.08 | 7.34 | 4.47 | 4.94 | 662.02 |
| TiNCl | 3.43 | 6.41 | 3.89 | 4.29 | 510.66 |
| TiNBr | 4.55 | 5.10 | 3.14 | 3.35 | 385.92 |
| TiNI | 5.31 | 4.61 | 2.73 | 3.03 | 333.84 |

The Debye temperature decreases with increasing atomic weights of the halogen atoms in TiN$X$. It is very interesting to note that $\theta_D$ versus inverse square root of halogen atomic mass curve show almost linear behavior for the compounds under study. This implies that the mass of the $X$ atomic species plays a prime role in determining the Debye frequency in these materials.

### 3.4 *Electronic band structure and density of states of* TiN$X$

Results of electronic band structure calculations with the optimized crystal structure of TiN$X$ are presented in this section. The electronic energy dispersion curves along the high-symmetry directions of the orthorhombic BZ are shown in Figs. 2. The calculated total and partial energy density of states (TDOSs and PDOSs, respectively), as a function of energy, ($E$), are presented in Figs. 3. The straight line denotes the Fermi level, $E_F$, which has been set to zero. To understand the contribution of each atomic orbital to the TDOSs, the PDOSs have been calculated for Ti, N, and $X$ atoms in TiN$X$.

Fig. 2a shows the electronic band structure for TiNF. A direct band gap of magnitude 0.579 eV is seen centered at the $\Gamma$ point of the BZ. The energy dispersion curves just above and below the Fermi energy are the ones responsible for charge transport and bonding properties. These important electronic energy bands show varying degree of dispersions. For example, the $E(k)$ curves close to $E_F$ along the $\Gamma$-$Z$ direction are non-dispersive, whereas the bands along the $Z$-$U$ direction in the momentum space show fairly dispersive character. The bands in the $X$-$S$ and $S$-$Y$ directions are moderately dispersive in the energy range from -2.0 to 2.0 eV. The low energy $E(k)$ curves along also show similar feature. Considering the geometry of the BZ for orthorhombic symmetry, the band structure features of TiNF reveal significant electronic anisotropy. The band structure is anisotropic both within the *ab*-plane and out-of-plane *c*-direction. Almost non-dispersive $E(k)$ in the *c*-direction implies that effective mass of the charge carriers are very high in this direction and charge transport should be dominated by the in-plane component in TiNF. Fig. 2b illustrates the electronic band structure of the TiNCl compound. Like TiNF, TiNCl is also a direct band gap semiconductor with a band gap of magnitude 0.503 eV centered at the $\Gamma$ point of the BZ. The band structure of TiNCl exhibits high degree of qualitative and quantitative similarities with that of TiNF, characterized by high degree of electronic anisotropy. The electronic band structure of TiNBr is presented in Fig. 2c. Once again the gross features of the $E(k)$ curves are quite similar to those found for TiNF and TiNCl. The magnitude of the direct band gap centered at the origin of the BZ for TiNBr is 0.482 eV. The deep lying bands below the Fermi level (-4eV and) and high energy bands above the Fermi level (4 eV and above) are quite dispersive in nature in all TiN$X$ compounds. Fig. 2d shows the electronic band structure of TiNI. This compound is characterized by extremely narrow band gap of 0.097 eV. The direct band gap has been shifted from the $\Gamma$ point to the $Z$ point in the BZ. The band structure though possesses anisotropy; its degree is lower than that found for TiNF, TiNCl, and TiNBr. Extremely low value of the band gap and relatively dispersive nature of the $E(k)$ curves of TiNI imply that this material is a weak semiconductor on the verge of metallicity.

To explore the relative contributions of different atomic orbitals to the TDOSs of TiN$X$, we have calculated the atom-resolved partial electronic density of states first. Figs. 3 represent the PDOSs and TDOSs for all the compounds under study. From the DOS profile for TiNF (Fig. 3a), it is seen that the high DOS part in the valence band (VB) in the energy range from 0.0 to -8.0 eV is mainly due to the contributions from the F 2p electronic orbitals with some contribution from the N 2s electronic states. Ti 3d electrons play a minor role. Significant band overlap between F 2p and N 2s orbitals in this energy range indicates strong bonding characteristics between these two atomic species. The DOS within the conduction band (CB) in the energy range from 0.50 eV to 7.50eV is dominated by the Ti 3d electronic states. There is a small contribution coming from the N 2s orbitals. It is noteworthy that the DOS in this energy range is splitted. This separates the bonding and antibonding electronic states in the CB. Next, we investigate the electronic energy density of states features of TiNCl as shown in Fig. 3b. The PDOS and TDOS features below the Fermi level in the VB of TiNCl are quite similar to those for TiNF. The only difference being Cl 3p states take part in the density of states instead of the 2p states for F. The DOS in the CB in the high energy part around 8.0 eV, on the other hand shows an extra peak derived completely from the Ti 3p electronic states. This feature is largely absent in the case of TiNF. The PDOS and TDOS profiles for TiNBr are shown in Fig. 3c. The DOS in the VB below the Fermi level is quite similar to those for TiNF and TiNCl. In the case of TiNBr, it is the Br 4p orbitals that are strongly hybridized with the N 2s electronic states to constitute the DOS structure in the energy range from 0.0 eV to -7.0 eV. The DOS feature in CB above $E_\text{F}$ is constituted from contributions primarily due to Ti 3d electronic states with almost equal contributions at ~ 8.0 eV from Br 4s and Ti 3p orbitals. At lower energy (2 – 6 eV) N 2s states are hybridized strongly with the Ti 3d electronic states. The DOS profile for TiNI is presented in Fig. 3d. The profile in the VB again shows features similar to the other compounds. In this case I 5p states are contributing to the TDOS in the VB together with the N 2s states. For energies above the Fermi level the sharp DOS feature at the ~ 2.5 eV is somewhat diminished in this compound. The TDOS in the CB is broadened and show shallow splitting for TiNI. Like the other three compounds, the TDOS in the CB arises from the Ti 3d, 3p, N 2s, and I 5s orbitals.

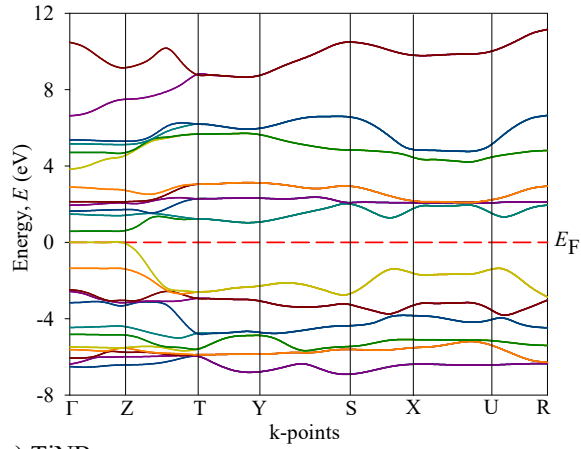
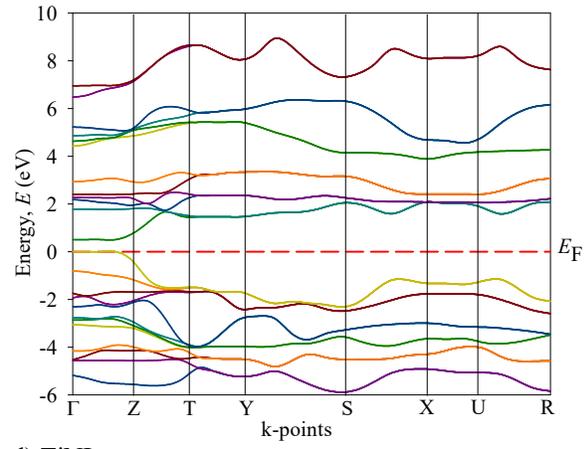
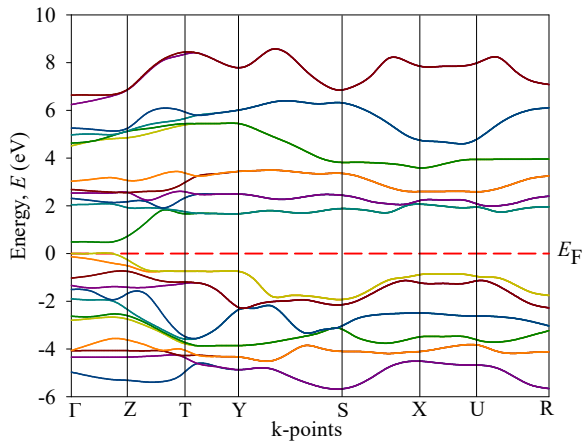
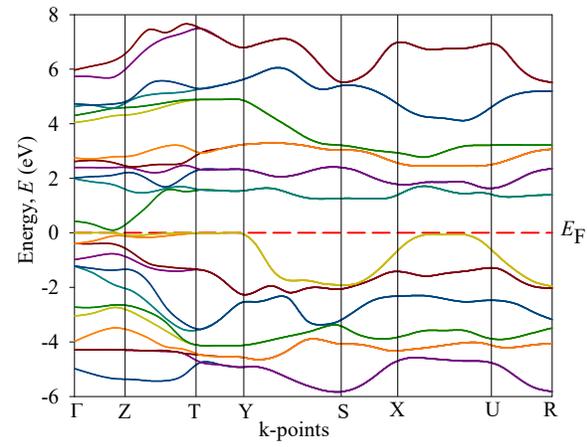

Figure 2: The band structure of Ti N*X*. Energy dispersion curves are shown along the high symmetry directions in the BZ. The Fermi level (marked by the dashed line) is set at zero.

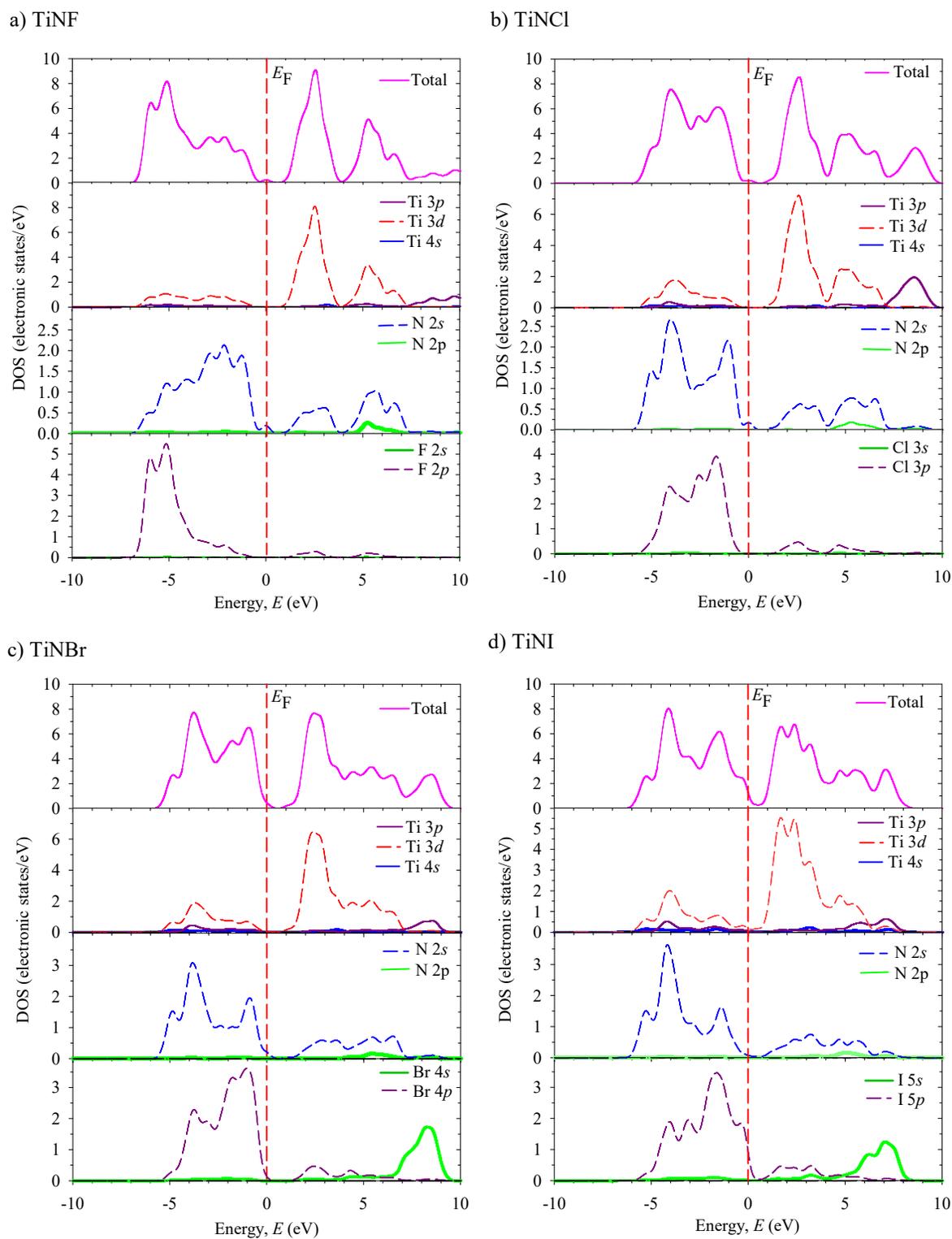

Figure 3: The PDOSs and the TODSs of TiN*X*. The Fermi level (marked by the dashed line) is set at zero.

## 3.5 Bond population analysis of TiNX

In order to explore the bonding nature of these layered ternaries in greater depth, the Mulliken bond populations [52] are investigated. The Hirshfeld population analysis [53] has also been performed. The main findings from the Mulliken population analysis (MPA) and Hirshfeld population analysis (HPA) are summarized briefly in this section. The results of these analyses are presented in Table 7.

**Table 7**
Charge spilling parameter (%), orbital charges (electron), atomic Mulliken charges (electron), and Hirshfeld charge (electron) in TiNX compounds.

| Compound | Atoms | Charge spilling (%) | s | p | d | Total | Mulliken charge | Hirshfeld charge |
|---|---|---|---|---|---|---|---|---|
| TiNF | Ti |  | 2.26 | 6.29 | 2.30 | 10.86 | 1.14 | 0.45 |
|  | N | 0.13 | 1.71 | 3.99 | 0.00 | 5.70 | -0.70 | -0.33 |
|  | F |  | 1.94 | 5.50 | 0.00 | 7.44 | -0.44 | -0.12 |
| TiNCl | Ti |  | 2.31 | 6.34 | 2.43 | 11.08 | 0.92 | 0.35 |
|  | N | 0.17 | 1.71 | 4.00 | 0.00 | 5.71 | -0.71 | -0.34 |
|  | Cl |  | 1.93 | 5.28 | 0.00 | 7.21 | -0.21 | -0.02 |
| TiNBr | Ti |  | 2.36 | 6.56 | 2.45 | 11.36 | 0.64 | 0.33 |
|  | N | 0.13 | 1.71 | 3.99 | 0.00 | 5.70 | -0.70 | -0.34 |
|  | Br |  | 1.79 | 5.15 | 0.00 | 6.94 | 0.06 | 0.01 |
| TiNI | Ti |  | 2.36 | 6.62 | 2.49 | 11.46 | 0.54 | 0.29 |
|  | N | 0.14 | 1.71 | 3.99 | 0.00 | 5.70 | -0.70 | -0.35 |
|  | I |  | 1.82 | 5.02 | 0.00 | 6.84 | 0.16 | 0.05 |

The Mulliken atomic charge density of Ti, N, and F in TiNF are 1.14, -0.70 and -0.44 electron, respectively. All these deviate from the normal value expected in a purely ionic state. This deviation partly reflects the covalent bonding characters among Ti, N and F atoms which form due to the hybridization among the Ti 3d, N 2s, and F 2p electronic orbitals as suggested by the PDOS profiles. Similar partial covalent bonding characters are seen from TiNCl, TiNBr and TiNI, consistent with respective PDOS features. It is interesting to note that as one progresses F to I, the magnitude of the Mulliken charge decreases systematically. This implies that the compounds are becoming more and more ionic as the electronegativity of the X atom increases. It is observed that the band spilling parameters very low for all the compounds. This results in almost zero TDOS at the Fermi level for TiNX as revealed by our electronic band structure calculations.

Because of the strong basis set dependence of MPA, it often gives results in contradiction to chemical intuition and overestimates the degree of covalency. For this reason, we have also employed HPA. This has practically no basis set dependence and can provide with a more physically meaningful result compared to the MPA. Table 7 shows the Hirshfeld charge. The general trend in the variation of the Hirshfeld charge for different *X* atoms follow the same qualitative feature as for the Mulliken charge. The magnitude of the Hirshfeld charge is smaller than that of Mulliken charge, a common feature often observed in other systems [54]. The gross features of the bonding nature among different atomic species in TiN*X* remain unchanged. The level of covalency and iconicity is lower in the HPA. This may imply that some metallic bondings are also present in TiN*X* compounds.

### 3.6 *Optical properties of* TiN*X*

Optical properties of a compound determine its electronic response to the incident electromagnetic radiation. The response to the infrared, visible and ultraviolet spectra is particularly important from the view of optoelectronic and photovoltaic device applications. This response to the incident electromagnetic wave is completely determined by the various energy (frequency) dependent optical constants, namely, the real and the imaginary parts of the dielectric constants, $\varepsilon_1(\omega)$ and $\varepsilon_2(\omega)$, respectively, real part of refractive index $n(\omega)$, extinction coefficient $k(\omega)$, the optical conductivity $\sigma(\omega)$, reflectivity $R(\omega)$, absorption coefficient $\alpha(\omega)$, and the loss function $L(\omega)$. The imaginary part of the dielectric constant was calculated using the electronic band structures obtained for TiN*X* to estimate the photon induced transitions between electronic states. It should be mentioned that the intraband electronic transitions are not very important for semiconductors. Such contribution to the optical properties affects mainly the low-energy part of the optical spectra in metallic compounds.

We have illustrated the calculated optical constants for TiN*X* in Figs. 4 – 7. The variation of the real and imaginary parts of the dielectric constant of TiNF is shown in Fig. 4a. The value of the static real part, $\varepsilon_1(0)$, is quite high, ~ 16. The real part of the dielectric constant is associated with the polarizability of the crystal which expresses the linear response of the material to an incident electromagnetic radiation. The value of the real part is directly related to the dipole inter-band transition probability. This transition is subject to selection rules and DOS of the states involved. It is seen that $\varepsilon_1(\omega)$ is high up to an energy ~ 2.0 eV. Above 2.0 eV it decreases sharply and crosses zero at ~ 3.0 eV. This feature leads to a peak both in the reflectivity and absorption spectra, implying non-propagating character of the incident electromagnetic waves with $\hbar\omega$ >3.0eV. The imaginary part of the dielectric function is intimately linked with the electronic band structure. It is seen from Fig. 4a that the rapid decrease in the $\varepsilon_1(\omega)$ above 2.0 eV is concomitant of a sharp increase in the imaginary part, $\varepsilon_2(\omega)$. This results in a peak in $\varepsilon_2(\omega)$ at ~ 3.0 eV. This peak arises due to photon induced electronic transitions between occupied and unoccupied states of F 2p, N 2s, and Ti 3d orbitals as revealed by the DOS spectrum. The broad feature seen in the energy range 5.0 – 8.0 eV also involves various optical transitions among the F 2p, N 2s, and Ti 3d orbitals at higher energies. The refractive index and extinction coefficient are plotted in Fig. 4b. The structure of $n(\omega)$ and $k(\omega)$ follows closely the energy dependences of $\varepsilon_1(\omega)$ and $\varepsilon_2(\omega)$, respectively, as expected. The refractive index in the visible range is quite high, ~ 4 for TiNF. The optical conductivity is shown in Fig. 4c. $\sigma(\omega)$ demonstrates clear semiconducting characteristics. The peak in $\sigma(\omega)$ at ~ 3.0 eV corresponds to the peak in $\varepsilon_2(\omega)$ and arises because of creation of electron hole pairs in the conduction and valence bands, respectively. There is one to one correspondence between other relative weak energy dependent features of $\varepsilon_2(\omega)$ and $\sigma(\omega)$, since both these parameters are dependent of photon induced transition among electronic states. The reflectivity spectrum is shown in Fig. 4d. $R(\omega)$ is characterized by two peaks; one ~ 3.0 eV and the second at ~ 18 eV. The second peak is sharper and is indicative of metallic reflection near the plasma oscillation edge. From low energy to ultraviolet region, $R(\omega)$ exhibits largely nonselective behavior and stays around 40%. Fig. 4e presents the optical absorption spectrum. The onset of absorption is ~ 0.60 eV, very close to the band gap value obtained from the electronic band structure calculations. The absorption coefficient is quite high in the infrared to ultraviolet region. There is a prominent peak at ~ 17 eV, which is due to the onset of plasma oscillation. The frequency dependent energy loss function is shown in Fig. 4f. The loss function of a compound is an important parameter in the optical study which is useful for understanding the screened charge excitation spectra, especially the collective excitations produced by a swift electron traversing a solid. The highest peak of the energy loss spectrum appears at a particular incident light frequency (energy) known as the bulk screened plasma frequency.

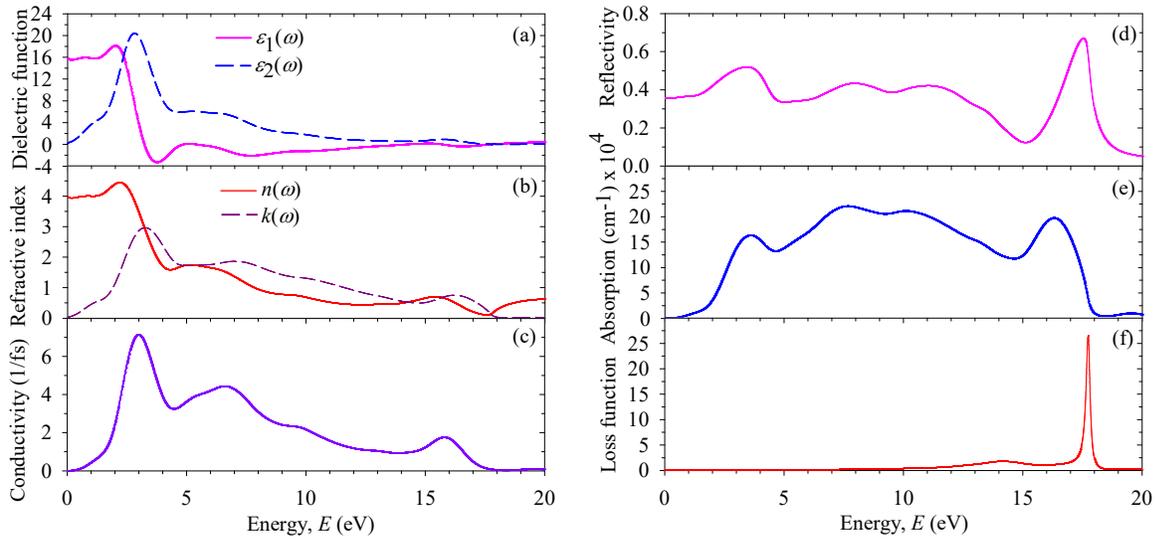

Figure 4: (a) $\varepsilon_1(\omega)$ and $\varepsilon_2(\omega)$, (b) $n(\omega)$ and $k(\omega)$, (c) $\sigma(\omega)$, (d) $R(\omega)$, (e) $\alpha(\omega)$, and (f) L$(\omega)$ of TiNF.

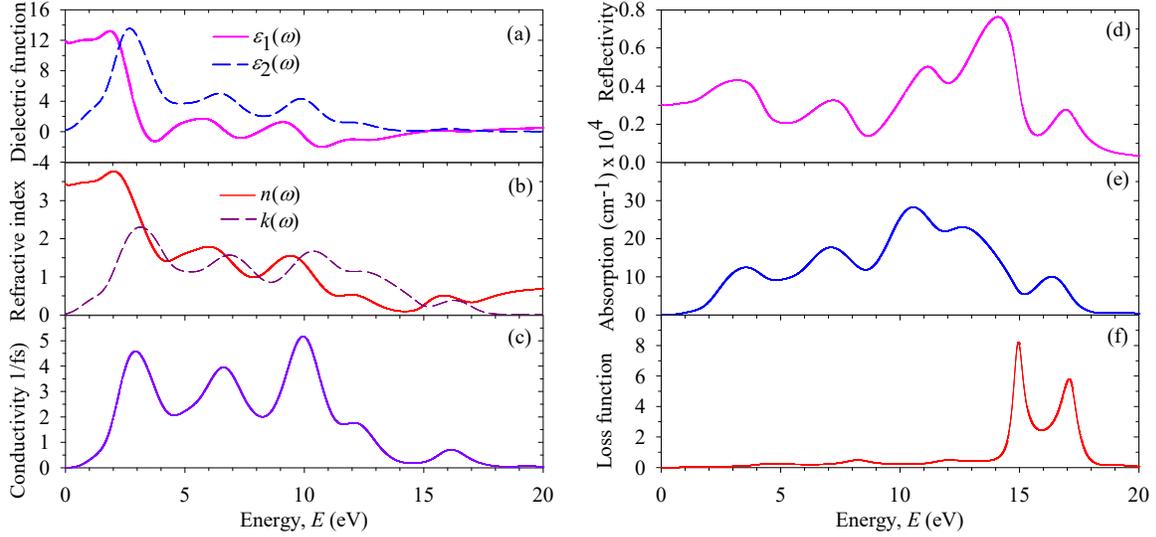

Figure 5: (a) $\varepsilon_1(\omega)$ and $\varepsilon_2(\omega)$, (b) $n(\omega)$ and $k(\omega)$, (c) $\sigma(\omega)$, (d) $R(\omega)$, (e) $\alpha(\omega)$, and (f) $L(\omega)$ of TiNCl.

The frequency dependences of the optical constants for TiNCl are shown in Figs. 5. The real and imaginary parts of the dielectric constants are shown in Fig. 5a. The gross features of dielectric constants are quite similar to those for TiNF. The static value of $\varepsilon_1$ is reduced, implying that TiNCl is less polarizable compared to TiNF. Additional low intensity peaks in the $\varepsilon_1(\omega)$ and $\varepsilon_2(\omega)$ arise due to the change in the electronic band structure with respect to TiNF. Instead of the F 2p orbitals in TiNF, Cl 3p orbitals contribute in case of TiNCl. The frequency dependent $n(\omega)$ and $k(\omega)$ are given in Fig. 5b. The frequency dependent refractive index and extinction coefficient show features as suggested by the dielectric constants. The refractive index of TiNCl is high ~ 3.5 over an extended region of the electromagnetic spectrum covering the visible and infrared region. Fig. 5c shows the optical conductivity. Vanishing optical conductivity at zero energy confirms the semiconducting nature of TiNCl. The peaks in the $\sigma(\omega)$ spectrum correspond to the peaks in the $\varepsilon_2(\omega)$ profile as expected. Fig. 5d illustrates the $R(\omega)$ spectrum. The reflectivity of TiNCl is slightly lower than that of TiNF in the energy range 0.0 – 10.0 eV. There is broad and high peak centered around 14 eV. This peak in the ultraviolet region is related to the plasma excitation. The energy dependent absorption coefficient of TiNCl is shown in Fig. 5e. The onset of absorption is at ~ 0.50 eV. This value agrees quite well with the band gap energy obtained from the band structure calculations. $\alpha(\omega)$ spectrum shows high values in the ultraviolet region in the energy range 10.0 – 13.0 eV. The energy loss spectrum is shown in Fig. 5f.

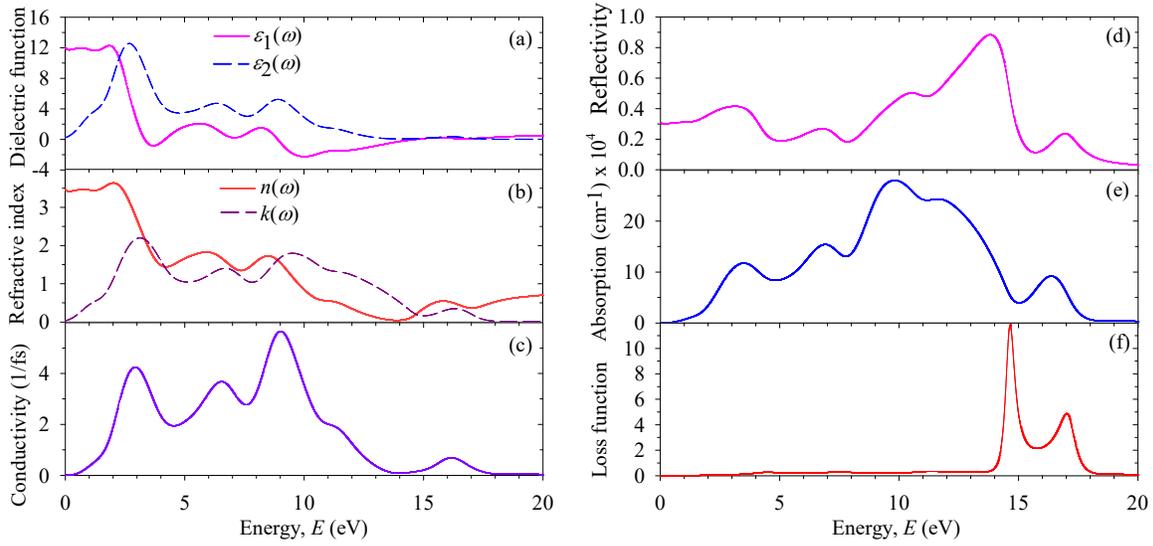

Figure 6: (a) $\varepsilon_1(\omega)$ and $\varepsilon_2(\omega)$, (b) $n(\omega)$ and $k(\omega)$, (c) $\sigma(\omega)$, (d) $R(\omega)$, (e) $\alpha(\omega)$, and (f) $L(\omega)$ of TiNBr.

The optical constants for TiNBr are presented in Figs. 6. Once again the frequency dependent real and imaginary parts of the dielectric constants (Fig. 6a) exhibit features consistent with band structure calculations energy dependent DOS of TiNBr. The refractive index and extinction coefficient are shown in Fig. 6b. Both the dielectric constants and the refractive index and extinction coefficient of TiNBr and TiNCl are quite similar both qualitatively and quantitatively. The refractive index of TiNBr is also high, ~ 3.5 in the energy range from 0.0 to 3.0 eV. $\sigma(\omega)$ clearly shows in Fig. 6c, semiconducting feature with a gap of around 0.40 eV. This completely agrees with the band structure calculations. The reflectivity spectrum is depicted in Fig. 6d. $R(\omega)$ is somewhat lower compared to TiNF and TiNCl in the energy range 0.0 – 10.0 eV. Above 10 eV it increases sharply and attains a maximum value ~ 85% at ~ 14.0 eV. This peak is concomitant to the peak in the loss function in TiNBr. The energy dependent absorption coefficient spectrum is shown in Fig. 6e. The loss function is shown in Fig. 6f. The absorption characteristics of TiNBr are quite similar to those for TiNCl.

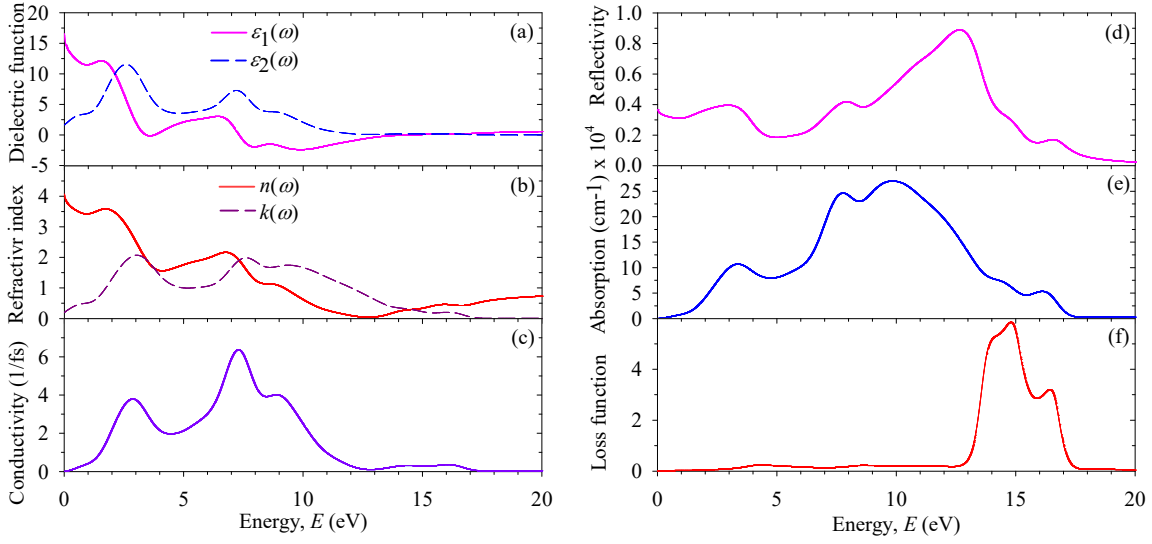

Figure 7: (a) $\varepsilon_1(\omega)$ and $\varepsilon_2(\omega)$, (b) $n(\omega)$ and $k(\omega)$, (c) $\sigma(\omega)$, (d) $R(\omega)$, (e) $\alpha(\omega)$, and (f) $L(\omega)$ of TiNI.

$\varepsilon_1(\omega)$ and $\varepsilon_2(\omega)$ spectra for TiNI is given in Fig. 7a. Like the other three compounds the structures in $\varepsilon_1(\omega)$ and $\varepsilon_2(\omega)$ complement the electronic band structure to a high degree. The frequency dependent $n(\omega)$ and $k(\omega)$ are shown in Fig. 7b. Unlike the cases in TiNF, TiNCl, and TiNBr, $n(\omega)$ for TiNI shows greater dispersion in the low energy region, with $n(0)$ approaching 4.0. The optical conductivity can be seen in Fig.7c. The optical conductivity shows an increase from almost zero energy. This is because this compound has a very small band gap compared to the other three. $R(\omega)$ is shown in Fig. 7d. The reflectivity in the infrared and visible region is quite low. It increases in the ultraviolet region and peaks at ~ 13.0 eV showing a characteristic qualitatively similar to other TiN$X$ compounds. The $\alpha(\omega)$ spectrum is shown in Fig. 7e. This compound shows strong absorption characteristics in the energy range from 7.0 – 12.0 eV. It is interesting to note that the magnitude of the optical conductivity tends to decrease slowly as one move from TiNF to TiNI. The loss spectrum is presented in Fig. 7f.

It is instructive to note that, even though the energy dependent optical parameters shown in this section was investigated within the LDA, the gross features remain almost identical if GGA is employed.

## 4 Discussion and conclusions

First-principles DFT based calculations have been carried out to investigate the structural, elastic, bonding, electronic, and optical properties layered ternary TiN$X$ semiconductors with orthorhombic symmetry. The elastic anisotropy, bonding characteristics and the energy dependent optical constants have been studied in detail for the first time. GGA, LDA, and hybrid HSE06 exchange-correlation potentials have been employed. LDA yields significantly better agreement with experimental parameters where available. The optimized lattice parameters agree quite well with earlier studies [13, 27]. The elastic constants and the moduli reveal that the compounds under study are relatively soft compared to many other layered ternaries [16 – 18, 26, 36 – 39]. All the compounds under study show significant anisotropy in their mechanical properties, characterized by the different anisotropy indices. TiNX are highly compressible when stress is applied along the *c*-direction. This is illustrated by the anisotropy factor for compressibility. This reflects strongly the layered structural and bonding features of TiN*X*. The machinability index of TiN*X* compares favorably to that for other layered ternaries [14, 16, 17]. All these compounds possess relatively high value of $\mu_M$, with the highest value for TiNI. This compound lies on the verge of metallicity and is quite soft elastically. It is quite interesting to note that almost all the elastic constants and moduli decrease systematically with increasing atomic number of the halogen atom. We presume that this behavior is probably linked to the systematic decrease in electronegativity down the group (I < Br < Cl < F). The variation in the electronegativity is expected to have a significant effect on the bonding nature and on the bonding strength. The variations of the Pugh's ratio and Poisson's ratio of TiN*X* with halogen atomic species are rather weak. This implies that the overall bonding character is not changed significantly as one moves from TiNF to TiNI. Both Pugh's ratio and the Poisson's ratio indicate that TiN*X* compounds are brittle in nature.

The Debye temperature of TiN*X* shows marked variation with *X*. For example, a high $\theta_D$ of 662.02 K for TiNF decreases systematically to 333.84 K for TiNI. This systematic variation and its correspondence to the atomic mass of the halogen atom indicate that the harmonic thermal vibrations of the *X* atoms mainly determine the $\theta_D$ of TiN*X*. This is interesting because $\theta_D$ determines a large number of thermal and charge transport characteristics of compounds. Therefore, it is possible to tune these properties in TiN*X*, via selecting a particular *X* species. High value of Debye temperature generally corresponds to stronger atomic bonding, phonon thermal conductivity, and melting temperature. This is the first systematic study of Debye temperature of TiN*X* to the best of our knowledge.

The electronic band structure calculations show direct band gap semiconducting nature of TiN*X*. The gap magnitudes for TiNF, TiNCl, TiNBr, and TiNI are 0.579 eV, 0.503 eV, 0.482 eV, and 0.097 eV, respectively. The magnitude of the gap decreases very gradually as one moves from TiNF to TiNBr. The gap value decreased drastically for TiNI. This material shows almost metallic character. A small variation in temperature will induce a large change in the electron and hole concentrations in this material. This property can be quite useful for device applications. The proximity of the top of the valence band to the Fermi level implies that hole type charge conduction should dominate in intrinsic TiN*X*. This proximity also implies that slight change in the *c*-axis lattice parameter or doping by intercalation between the atomic layers can drive this system to metallic regime. This probably leads to superconductivity in intercalated Ti*NX* where insertion of atomic layers changes both *c*-axis lattice parameter and adds charge carriers in the Ti*NX* system. Significant band overlaps between the F 2p, Cl 3p, Br 4p, I 5p and N 2s orbitals are found in TiNF, TiNCl, TiNBr, and TiNI, respectively in the high energy part of the VB. This hybridization indicates strong bonding tendency between these atoms. The electronic DOS within the CB is dominated by the Ti 3d electronic states in all the Ti*NX* compounds. The electronic band structure exhibits significant electronic anisotropy. The out-of-plane $E(k)$ features are weakly dispersive compared to the in-plane ones. This signifies anisotropic effective mass of the charge carriers. The structural anisotropy is concomitant to the electronic anisotropy. Woodword and Vogt [4] have calculated the electronic band structure of *MNX* (*M* = Zr, Ti; *X* = Cl, Br, I) using the extended Huckel method. The band structure features and the values of the band gaps obtained for TiNBr and TiNI in this study agree quite well with those obtained in Ref. [4]. But the theoretical band gap for TiNCl was 1.70 eV in Ref. [4] which is very large compared to the band gap of 0.503 eV obtained here. The estimated value in this study is in good agreement with the theoretical one (0.601 eV) given in Ref. [28]. It is perhaps suggestive to note that the band gap energy obtained by Woodword and Vogt for TiNCl resembles more to the typical *β* phase *MNX* layered compounds. Band gap values obtained in this study show excellent agreement with those found in Ref. [1] calculated via the Perdew–Burke–Ernzerhof (PBE) [55] exchange-correlation functional scheme.

The bonding nature was also explored via MPA and HPA. Both MPA and HPA indicate that covalent and ionic bonds dominate in Ti*NX*. Bond population analysis supports the hybridization features seen in the DOS of Ti*NX*.

The optical parameters of these layered ternaries have been studied for the first time in details. The peaks and structures in the real and imaginary parts of the dielectric constant show clear correspondence to the electronic band structure. Both optical conductivity and the absorption coefficient show clear semiconducting behavior. The magnitude of the optical band gaps agree to those obtained from the band structure. The reflectivity of Ti*NX* compounds is low over a wide range of energy from infrared to near ultraviolet region. It decreases with the increase of the atomic mass of *X*. All these compounds show relatively high absorption in the ultraviolet region. The reflectivity is also quite high in Ti*NX* compounds in the mid-ultraviolet in an energy range from 12 to 14 eV. One interesting optical characteristic of Ti*NX* is its high refractive index in the infrared and visible range. This is important because high refractive index materials are useful as anti-reflection coating and also in various optoelectronic devices like the light emitting diode (LED).

To summarize, structural, elastic, electronic, bonding, and optical properties of Ti*NX* semiconductors have been studied in this paper. These versatile layered compounds show a number of attractive features which can be tuned via the halogen atom. We hope that this study will inspire both the theorists and the experimentalists to investigate these compounds in greater details in future.

**List of references**

1. Yan Liang, Ying Dai, Yandong Ma, Lin Ju, Wei Wei, Baibiao Huang, Novel titanium nitride halide TiNX (X = F, Cl, Br) monolayers: potential materials for highly efficient excitonic solar cells, J. Mater. Chem. A (2018) DOI: 10.1039/c7ta09662c.
2. J. Liu, X.-B.Li, D. Wang, H. Liu, P. Peng, L.-M.Liu, Single-layer Group-IVB nitride halides as promising photocatalysts, J. Mater. Chem. A 2 (2014) 6755–6761.
3. R. Juza, J. Heners, ÜberNitridhalogenide des Titans und Zirkons, Z. Anorg. Allg. Chem. 332 (1964) 159–172.
4. P.M. Woodward, T. Vogt, Electronic Band Structure Calculations of the *M*N*X* (*M* = Zr, Ti; *X* = Cl, Br, I) System and Its Superconducting Member, Li-Doped *β*-ZrNCl, Journal of Solid State Chemistry 138 (1998) 207-219.
5. Christian M. Schurz, Larysa Shlyk, Thomas Schleid, Rainer Ni, Superconducting nitride halides MNX (M = Ti, Zr, Hf; X = Cl, Br, I), Z. Kristallogr. 226 (2011) 395–416.
6. Shoji Yamanaka, High-$T_c$ Superconductivity in Electron-Doped Layer Structured Nitrides, Annual Review of Materials Science 30 (2000) 53-82 and references therein.
7. Shuai Zhang, Masashi Tanaka, ErikaWatanabe, Haikui Zhu, Kei Inumaru, Shoji Yamanaka, Superconductivity of alkali metal intercalated TiNBr with α-typenitride layers, Supercond. Sci. Technol. 26 (2013) 122001.
8. Shuai Zhang, Masashi Tanaka, Haikui Zhu, Shoji Yamanaka, Superconductivity of layered β-HfNCl with varying electron-doping concentrations and interlayer spacings, Supercond. Sci. Technol. 26 (2013) 085015.
9. G. J. Snyder, E. S. Toberer, Complex thermoelectric materials, Nature materials 7 (2008) 105-114.
10. Shuofeng Zhang, Ben Xu, Yuanhua Lin, Cewen Nan, Wei Liu, First-principles study of the layered thermoelectric material TiNBr, arXiv:1710.02495v1.
11. Quan Yin, Erik R.Ylvisaker, Warren E. Pickett, Spin and charge fluctuations in α-structure layered nitride superconductors, Phys. Rev. B83 (2011) 014509.
12. Shoji Yamanaka, Keita Umemoto, Superconductivity of TiNCl intercalated with diamines, Physica C 470 (2010) S693.
13. B. Altintas, STRUCTURAL AND ELECTRONIC PROPERTIESOF *α*-TiNX (X: F,Cl, Br, I). AN *AB INITIO* STUDY, Journal of Theoretical and Computational Chemistry 10 (2011) 65-74) and references therein.
14. F. Parvin, S.H. Naqib, Structural, elastic, electronic, thermodynamic, and optical properties of layered BaPd$_2$As$_2$ pnictide superconductor: a first principles investigation, Journal of Alloys and Compounds 780 (2019) 452.

15. A. Chowdhury, M.A. Ali, M.M. Hossain, M.M. Uddin, S.H. Naqib, A.K.M.A. Islam, Predicted MAX Phase $Sc_2InC$: Dynamical Stability, Vibrational and Optical Properties, Physica Status Solidi: B (2017) DOI: 10.1002/pssb.201700235.
16. F.P. Parvin, S.H. Naqib, Elastic, electronic, and optical properties of recently discovered superconducting transition metal boride NbRuB: an ab-inito investigation, Chin. Phys. B 26 (2017) 106201.
17. M.A. Ali, M.A. Hadi, M.M. Hossain, S.H. Naqib, A.K.M.A. Islam, First-principles calculations of structural, elastic and electronic properties of MoAlB, Physica Status Solidi: B (2017) DOI: 10.1002/pssb.201700010.
18. M. Roknuzzaman, M.A. Hadi, M.J. Abden, M.T. Nasir, A.K.M.A. Islam, M.S. Ali, K. Ostrikov, S.H. Naqib, Physical properties of predicted $Ti_2CdN$ versus existing $Ti_2CdC$ MAX phase: An ab-initio study, Computational Materials Science 113 (2016) 148.
19. W. Kohn, L. J. Sham, Self-consistent equations including exchange and correlation effects, Phys. Rev. 140 (1965) A1133.
20. S. J. Clark, M. D. Segall, C. J. Pickard, P. J. Hasnip, M. I. J. Probert, K. Refson, M. C. Payne, First principles methods using CASTEP, Z. Kristallographie 220 (2005) 567-570.
21. T. H. Fischer, J. Almlof, General methods for geometry and wave function optimization, J. Phys. Chem. 96 (1992) 9768.
22. H. J. Monkhorst, J. D. Pack, Special points for Brillouin-zone integrations, Phys. Rev. B 13 (1976) 5188.
23. Fazle Subhan, SikanderAzam, Gulzar Khan, Muhammad Irfan, Shabbir Muhammad, Abdullah G. Al-Sehemi, S.H. Naqib, R. Khenata, SaleemAyaz Khan, I.V. Kityk, Bin Amin, Elastic and optoelectronic properties of $CaTa_2O_6$ compounds: Cubic and orthorhombic phases, Journal of Alloys and Compounds (2019) DOI: 10.1016/j.jallcom.2019.01.140.
24. M.A. Ali, M.T. Nasir, M.R. Khatun, A.K.M.A. Islam, S.H. Naqib, Ab initio investigation of vibrational, thermodynamic, and optical properties of $Sc_2AlC$ MAX compound, Chinese Physics B 25 (2016) 103100.
25. M.A. Ali, M. Roknuzzaman, M.T. Nasir, A.K.M.A. Islam, S.H. Naqib, Structural, elastic, electronic and optical properties of $Cu_3MTe_4$ (M = Nb, Ta) sulvanites: An ab-initio study, Int. J. Mod. Phys. B (2016) DOI: 10.1142/S0217979216500892.
26. M.A. Hadi, M.S. Ali, S.H. Naqib, A.K.M.A. Islam, Band structure, Hardness, thermodynamic, and optical properties of superconducting $Nb_2AsC$, $Nb_2InC$, and $Mo_2GaC$ from first-principles, Int. J. Comp. Mat. Sci. Eng. 02 (2013) 1350007.
27. D. A. Headspith, and G. M.Francesconi, Transition metal pnictide-halides: A class of under-explored compounds, Topics in Catalysis 52(11) (2009) 1611-1627.
28. Lukas Muechler, Leslie M. Schoop, Claudia Felser, On the superconductivity of TiNCl and ZrNCl: Alocal bonding perspective, arXiv:1408.3099v1.
29. https://chem.libretexts.org/Bookshelves/Inorganic_Chemistry

30. M.A. Hadi, N. Kelaidis, S.H. Naqib, A. Chroneos, A.K.M.A. Islam, Mechanical behaviors, lattice thermal conductivity and vibrational properties of a new MAX phase $Lu_2SnC$, J. Phys. Chem. Solids 129 (2019) 162.
31. P. Ravindran, Lars Fast, P. A. Korzhavyi, B. Johansson, J. Wills, O. Eriksson, Density functional theory for calculation of elastic properties of orthorhombic crystals: Application to $TiSi_2$, J. Appl. Phys. 84 (1998) 4891.
32. R. Hill, The Elastic Behaviour of a Crystalline Aggregate, Proc. Phys. Soc. London 65 (1952) 350.
33. S.F. Pugh, Relations between the elastic moduli and the plastic properties of polycrystalline pure metals, Philos. Mag. 45 (1954) 823-843.
34. W. Voigt, Lehrbuch der Kristallphysik (Taubner, Leipzig, 1928).
35. A. Reuss, Berechnung der Fließgrenze von Mischkristallen auf Grund der Plastizitätsbedingungfür Einkristalle, Z. Angew. Math. Mech. 9 (1929) 55.
36. P. Barua, M.M. Hossain, M.A. Ali, M.M. Uddin, S.H. Naqib, A.K.M.A. Islam, Effects of transition metals on physical properties of $M_2BC$ (M = V, Nb, Mo and Ta): a DFT calculation, Journal of Alloys and Compounds 770 (2019) 523.
37. F. Sultana, M.M. Uddin, M.A. Ali, M.M. Hossain, S.H. Naqib, A.K.M.A. Islam, First principles study of $M_2InC$ (M = Zr, Hf and Ta) MAX phases: The effect of M atomic species, Results in Physics 11 (2018) 869.
38. M.A. Ali, M.M. Hossain, M.A. Hossain, M.T. Nasir, M.M. Uddin, M.Z. Hasan, A.K.M.A. Islam, S.H. Naqib, Recently synthesized $(Zr_{1-x}Ti_x)_2AlC$ ($0 \leq x \leq 1$) solid solutions: Theoretical study of the effects of M mixing on physical properties, Journal of Alloys and Compounds 743 (2018) 146.
39. M.A. Hadi, M. Roknuzzaman, A. Chroneos, S.H. Naqib, A.K.M.A. Islam, R.V. Vovk, K. Ostrikov, Elastic and Thermodynamic Properties of $(Zr_{3-x}Ti_x)AlC_2$ MAX-Phase Solid Solutions, Computational Materials Science 137 (2017) 318.
40. M. H. Ledbetter in *Materials at Low Temperatures*, edited by R. P. Reed and A. F. Clark, American Society for Metals 1983.
41. G. N. Greaves, A. L. Greer, R. S. Lakes, T. Rouxel, Poisson's ratio and modern materials, NATURE MATERIALS 10 (2011) 823 and references therein.
42. Christopher M. Kube, Elastic anisotropy of crystals, AIP ADVANCES 6 (2016) 095209.
43. Shivakumar, Ranganathan, M. Ostoja-Starzewski, Universal Elastic Anisotropy Index, Phys. Rev. Lett.101 (2008) 055504.
44. D.H. Chung and W.R. Buessem, in Anisotropy in *Single Crystal Refractory Compound*, Edited by F. W. Vahldiek and S. A. Mersol, Vol. 2 (Plenum, New York, 1968).
45. M. Born, K. Huang, Dynamical Theory of Crystal Lattices (Oxford University Press, UK, 1998).
46. Felix Mouhat, Francois-Xavier Coudert, Necessary and sufficient elastic stability conditions in various crystal systems, Phys. Rev. B 90 (2014) 224104.

47. Z. Sun, D. Music, R. Ahuja, J. M. Schneider, Theoretical investigation of the bonding and elastic properties of nanolayered ternary nitrides, Phys. Rev. B 71 (2005) 193402.
48. O. L. Anderson and H. H. Demarest Jr., Elastic constants of the central force model for cubic structures: Polycrystalline aggregates and instabilities, J. Geophys. Res. 76 (1971) 1349.
49. M.A. Hadi, M.A. Alam, M. Roknuzzaman, M.T. Nasir, A.K.M.A. Islam, S.H. Naqib, Structural, elastic, and electronic properties of recently discovered ternary silicide superconductor $Li_2IrSi_3$: An *ab-initio* study, Chinese Physics B 24(11) (2015) 117401.
50. M.A. Hadi, S.H. Naqib, S.R.G. Christopoulos, A. Chroneos, A.K.M.A. Islam, Mechanical behavior, bonding nature and defect processes of $Mo_2ScAlC_2$: A new ordered MAX phase, Journal of Alloys and Compounds 724 (2017) 1167.
51. M.A. Alam, M.A. Hadi, M.T. Nasir, M. Roknuzzaman, F. Parvin, M.A.K. Zilani, A.K.M.A. Islam, S.H. Naqib, Structural, elastic, and electronic properties of newly discovered $Li_2PtSi_3$ superconductor: effect of transition metals, J. Supercond. Nov. Magn. (2016) DOI: 10.1007/s10948-016-3619-7.
52. R.S. Mulliken, Electronic Population Analysis on LCAO–MO Molecular Wave Functions. I., J. Chem. Phys. 23 (1955) 1833.
53. F.L. Hirshfeld, Bonded-atom fragments for describing molecular charge densities, Theor. Chim. Acta 44 (1977) 129.
54. M.I. Naher, F. Parvin, A.K.M.A. Islam, S.H. Naqib, Physical properties of niobium based intermetallics ($Nb_3B$; B = Os, Pt, Au): a DFT based ab-initio study, The European Physical Journal B 91 (2018) 289.
55. J. P. Perdew, A. Zunger, Self-interaction correction to density-functional approximations for many-electron systems, Phys. Rev. B 23 (1981) 5048.